\documentclass[prl,twocolumn]{revtex4}
\pdfoutput=1
\usepackage{amssymb,latexsym}
\usepackage{amsmath,amsbsy,bbm}
\usepackage{ifpdf}
\usepackage{epsfig,bm}
\usepackage{graphicx,comment}
\usepackage{color}
\usepackage{soul}
\usepackage{mathtools}
\usepackage{comment}
\usepackage[normalem]{ulem}
\begin{document}

\title{Unexpected results of the phase transitions of four-state Potts model on the square and the honeycomb lattices}
\author{Jhao-Hong Peng}
\affiliation{Department of Physics, National Taiwan Normal University,
88, Sec.4, Ting-Chou Rd., Taipei 116, Taiwan}
\affiliation{Department of Physics, Box 90305, Duke University, Durham, NC 27708, USA}
\author{Fu-Jiun Jiang}
\email[]{fjjiang@ntnu.edu.tw}
\affiliation{Department of Physics, National Taiwan Normal University,
88, Sec.4, Ting-Chou Rd., Taipei 116, Taiwan}

\begin{abstract}

It is widely believed that the phase transition for the four-state ferromagnetic Potts model on the 
square lattice is of the pseudo-first order.
Specifically, it is expected that first-order phase transition behavior is found on small
lattices and that the true nature of second-order phase transition only emerges with large system sizes. 
It is also intuitively expected that for other geometries, the types of the associated phase transitions should 
be identical to that of the square lattice. 
However, after simulating more than 16 million spins for the four-state Pott model, 
we observe that a feature of first-order phase transition persists on the square lattice.
Additionally, a characteristic of second-order phase transition already appears on a small honeycomb lattice.
Indications of a pseudo-first-order phase transition were not found in our investigation. This suggests
that a thorough analytic calculation may be required to develop a better understanding of the presented results.
  
\end{abstract}

\maketitle

\section{Introduction}\vskip-0.3cm

The $q$-state (ferromagnetic) Potts model is a spin model that serves as a
generalization of the Ising model \cite{Pot52,Wu82}. Despite its 
simplicity, the $q$-state Potts model is well-suited for studying phase
transitions. Several analytic and numerical calculations are consequently
conducted for this model and its variations (see Ref.~\cite{Wu82} and
the references therein). Generalizations and extensions
of the Potts model are also applied to explore certain phenomena of high energy physics and foam \cite{Gra92}. 

It is known theoretically that the phase transition of $q$-state Potts
model is second order for $1\le q\le 4$ and first order for
$q \ge 5$ \cite{Wu82,Bax73,Bax78,Hin78}. It has also been established that
the critical temperature $T_c(q)$ for a given $q$ is calculated as 
$1/\log\left(1+\sqrt{q}\right)$. Because Potts models are thoroughly
studied in the literature, they are perfectly suited for exploring new
ideas, such as the machine learning phases of certain physical
systems \cite{Li18,Tan20}

It should be pointed out that although Potts models have been investigated in great detail both theoretically 
and numerically, some analytic predictions still need to be confirmed. For instance, while it is 
widely believed that the phase transition of the five-state Potts model on the square lattice is weakly first 
order, there is a lack of concrete and convincing numerical evidence establishing such. The correlation length of 
the phase transition of the five-state Potts model on the square lattice is estimated to be over 2000 \cite{Buf93}, and 
simulating systems with more than 4 million spins, as the one does in Ref.~\cite{Tse21}, is thus required to provide 
solid evidence to support this theoretical prediction.

The four-state Potts model on the square lattice is of theoretical interest.
In addition to receiving certain logarithmic corrections \cite{Car86,Sal97}, it has an unusual phase transition. 
Specifically, it has a pseudo-first-order phase transition: its transition is of the first-order on small system sizes, 
and its true nature of the second-order phase transition only emerges when the system size is sufficiently large.

Although there are numerical simulations of the four-state Potts model on 
the square lattice available, there is a lack of solid evidence showing that the associated transition is indeed a pseudo-first-order one. In addition, the question of whether the pseudo-first-order phase transitions also appears
for other geometries such as the honeycomb lattice has not been explored in detail. We thus opted to carry out a detailed investigation of the
phase transitions of the four-state Potts model using Monte Carlo simulations on both the square and the honeycomb lattices in this study.

Surprisingly, we observe that a feature of first-order phase transition persists even for a system size of more than 16 million spins on the square lattice. 
In particular, the first-order phase transition signal becomes stronger when the linear box size $L$ increases from a small one to a large one. 
A characteristic of second-order phase transition also already appears on a small honeycomb lattice. In other words, we did not observe any signals of 
pseudo-first-order phase transitions in our study. The results shown here imply that one may need to re-examine the associated analytic calculations.

The rest of the paper is organized as follows. After the introduction, we present
the model and the numerical outcomes in Sec. II and
Sec. III, respectively. We then conclude our investigation in Sec. IV. 

\begin{figure}
  \vskip-0.5cm

       \includegraphics[width=0.2\textwidth]{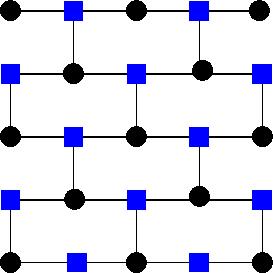}        
        \vskip-0.2cm
        \caption{The periodic 4 by 4 honeycomb lattice implemented in this
          study.}
        \label{honey}
\end{figure}
  
\section{The considered model and observable}

The Hamiltonian of the four-state Potts model considered here has the following expression \cite{Wu82}
\begin{equation}
\beta H_{\text{Potts}} = -\beta \sum_{\left< ij\right>} \delta_{\sigma_i,\sigma_j}.
\label{eqn}
\end{equation}
Here $\delta$ refers to the Kronecker function and the Potts variable
$\sigma_i$ at each site $i$ takes an integer value from $\{1,2,3,4\}$. 
The physical quantity investigated in this study is the energy
density $E$.

The periodic honeycomb lattice implemented in this study is depicted in
fig.~\ref{honey}.

\section{Numerical Results}

To study the nature of the phase transitions of the four-state Potts model on both the square and the honeycomb lattices, we have carried out large-scale Monte Carlo calculations (MC) using the Wolff algorithm \cite{Wol89}. The data points are recorded once after every 10 MC sweeps. Since the established $T_c$ for the four-state Potts models on the square and the honeycomb lattices are given by 0.91024 and 0.62133 \cite{Wu82,Sal97,Yau22}, respectively, the simulations carried out here will focus on the vicinity of these two temperatures. For large lattices, the associated results are
obtained based on a few independent simulations.

One of the most considered (and useful) quantities to determine the nature of a phase transition is the histogram of energy density $E$. In particular, a two-peak structure with certain properties will appear for a first-order phase transition. Specifically, if one defines $\Delta F(L)$ to be the difference between the peak and the valley of the probability distribution of $E$ on a $L$ by $L$ square area, then theoretically $ln \left(\Delta F(L)\right) \propto L$ for a first-order phase transition in two-dimension. For a second-order phase transition, the mentioned quantity will saturate to a constant with $L$. Here we would like to point out that the described scenario for the first-order phase transition will only appear for those of a certain strength \cite{Bil95}. Hence, practically a phase transition is considered to be first (second) order if $ln\left(\Delta F (L)\right)$ grows (approaches a constant) with $L$. We will use this rule in this study.

\begin{figure*}
  \vskip-0.5cm
  
  \begin{hbox}
      {~~~~~~~~~~~
       \includegraphics[width=0.4\textwidth]{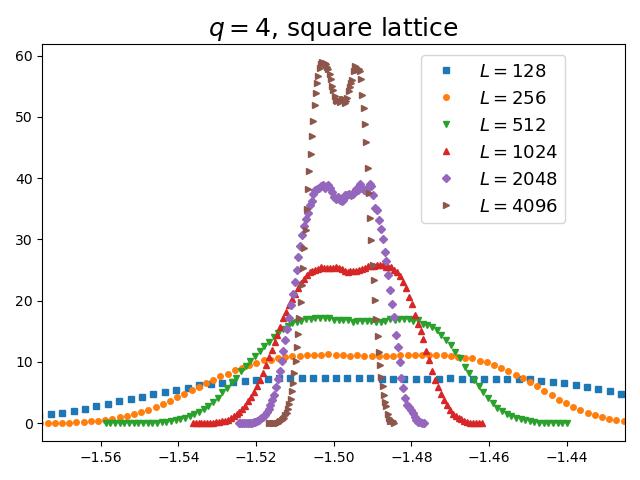}~~~~~~~
       \includegraphics[width=0.4\textwidth]{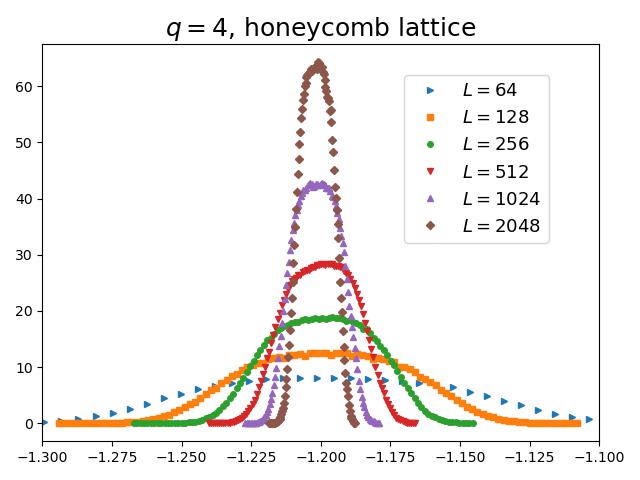} }
  \end{hbox}         
        \vskip-0.2cm
        \caption{The histograms of $E$ for various system linear sizes $L$
          of the four-state Potts model on the square (left panel) and the
          honeycomb (right panel) lattices. The results are obtained using the
        histogram function of pylab with bins = 100 or 50.}
        \label{square}
\end{figure*}

The histograms of $E$ for various system sizes ($L=128$, 256, 512, 1024, 2048, 4096) for the four-state Potts model on the square lattice are shown in the left panel of fig.~\ref{square}. As can be seen from the figure, the two-peak structure still shows up even for $L=4096$. Moreover, the
associated $\Delta F(L)$ seems to increase with $L$. This is rather unexpected since such a phenomenon belongs to a feature of first-order phase transition. Assuming the two peaks structure indeed eventually disappears, it is surprising that the indication of a first-order phase transition for the four-state Potts model (on the square lattice) persists even for a large system size of $L=4096$. Currently, it is beyond the scope of our study to simulate larger lattices than $L=4096$.

The histograms of $E$ for various system sizes ($L=64$, 128, 256, 512, 1024, 2048) for the four-state Potts model on the honeycomb lattice are shown in the right panel of fig.~\ref{square}. No two-peak strucure is observed even for $L=64$. In other words, the signal of second-order phase transition is already observed on small honeycomb lattices. This implies that the phase transition is truly second order, which diverges greatly from the case of the square lattice.

Table I lists the simulated temperatures for various box sizes for the
four-state Potts model on both the square and the honeycomb lattices.
Readers may reproduce fig.~\ref{square} based on the table.

\begin{table}
\begin{center}
\begin{tabular}{ |c|c|c| } 
  \hline
 $L$ &$T_c(L)$ & \text{lattice} \\
  \hline
 128 & 0.911130 & square\\ 
 256 &  0.910582& square \\ 
 512 & 0.910372 & square \\
 1024 & 0.910290 & square\\ 
 2048 & 0.910257 & square \\
 4096 & 0.9102455 & square \\
 64 & 0.6226 & honeycomb\\ 
 128 & 0.621860 & honeycomb \\
 256 & 0.621530 & honeycomb\\ 
 512 & 0.621415 & honeycomb \\ 
 1024 & 0.621362 & honeycomb \\
 2048 & 0.6213455 & honeycomb \\
 \hline
\end{tabular}
\caption{The temperatures used to produce the outcomes shown in
  fig.~\ref{square}.}
\end{center}
\end{table}

\section{Discussions and Conclusions}

In this study, we investigate the nature of the phase transitions of the four-state Potts model on both the square and the honeycomb lattices using the MC approach. Theoretically and intuitively, it is anticipated that both considered phase transitions should be of the pseudo-first order.

The outcomes shown here are rather unexpected. On the square lattice the signal of first-order transition persists even for very large system sizes, and, for the honeycomb lattice, characteristics of second-order phase transition appear earlier than anticipated on small lattices. Features of pseudo-first-order phase transition are never observed in our investigation.

In Refs.~\cite{Jin12,Jin13}, the phase transition of the four-state Potts model on the square lattice is studied. It is claimed in these references that the investigated phase transition is pseudo-first order based on the associated data up to $L=512$. In particular, it is found in Ref. \cite{Jin13} that for the probability distribution of the energy density, the dip between the two peaks does not increase appreciatively as one moves from $L=128$ to $L=512$. Although this is consistent with the observation obtained here, the dip increases significantly with $L$ once the system sizes are beyond $L=512$. This can be seen clearly from the left panel of fig.~\ref{square}.

It is interesting to note that in Ref.~\cite{Kal12}, which is a study of
the frustrated Ising model on the square lattice, the signal of first-order phase transition disappears only for $L \ge 1500$. If the same scenario indeed occurs for the four-state Potts model, then simulation of gigantic lattices is required to determine the truth of such.

In Ref.~\cite{Liu22}, the Baxter-Wu model, which is in the same universality class as the four-state Potts model, is studied. This investigation revealed that the latent heat decreases as the lattice size increases (The largest $L$ used in Ref.~\cite{Liu22} is $L=90$). This implies that the first-order phase transition signal becomes weaker with $L$. For the four-state Potts model on the square lattice, it is then certain that the characteristic of first-order phase transition gets stronger with $L$, at least for the system sizes considered in this study.

Based on the presented numerical results, a thorough theoretical calculation may be necessary to obtain a better understanding of the findings shown here.

Finally, it will be interesting to conduct similar studies for 2D four-state Potts model on the triangular and the kagome lattices.

{\it Note}: The four-state Potts model on the honeycomb lattice was
studied in Ref.~\cite{Mur19} using the Wang-Landau algorithm \cite{Wan01}. The result claimed in Ref.~\cite{Mur19} is different from that presented here.

\section*{Acknowledgement}\vskip-0.3cm
Partial support from National Science and Technology Council (NSTC) of Taiwan is
acknowledged.

\end{document}